# Efficient Asymmetric photothermal source: Designing a heating Janus-Nanojet

Javier González-Colsa[a], Alfredo Franco Pérez[a], Fernando Bresme[b], Fernando Moreno[a] and Pablo Albella[a*]

[a]Group of Optics, Department of Applied Physics, University of Cantabria, 39005, Santander, Spain.
[b]Department of Chemistry, Molecular Sciences Research Hub, Imperial College London, W12 0BZ, London, United Kingdom.

*Email: pablo.albella@unican.es

**Abstract:** Janus particles have flourished as subject of intensive research due to their synergetic properties and their promising use in different fields, especially in biomedicine. The combination of materials with radically different physical properties in the same nanostructure gives rise to the so-called Janus effects, allowing phenomena of a contrasting nature to occur in the same architecture. In particular, interesting advantages can be taken from a thermal Janus effect for photoinduced hyperthermia cancer therapies. Such therapies still have limitations associated to the heating control in terms of temperature stability and energy management. While previous studies have shown that some plasmonic single-material nanoparticles are somehow effective at killing cancer cells, it is necessary to investigate alternative plasmonic Janus nanoheaters to improve the heating efficiency and thermal control, mainly because the widespread single-material nanoheaters are highly homogeneous sources of heat, which implies that the surrounding biological medium is isotropically heated, equally affecting cancerous and healthy cells. A detailed thermoplasmonic study of the thermal Janus effect is still missing. Here we perform such study and demonstrate that doughnut-based Janus nanoparticles exhibit an outstanding photothermal control under practical illumination conditions, i.e., unpolarized light. Furthermore, we present novel and effective Janus nanoparticle designs that possess superior photothermal conversion features and unique directional heating capacity, being able to channel up to 91% of the total thermal energy onto a target. We discuss the implications of these innovative nanoparticles with regards to thermoplasmonics hyperthermia cancer therapy.

## 1. Introduction

Cancer is one of the major causes of mortality worldwide [1]. Curing cancer is a key priority to society. Scientists have developed a variety of approaches to fight cancer [2]. Radiation therapy, chemotherapy or surgery are regarded as the main clinical treatments but they suffer from many limitations, such as surgical complications and the serious side effects associated to radio and chemotherapy. Moreover, the efficacy of these three treatments is masked by metastasis phenomena [3] which render cancer cells hard to be completely eradicated, leading to a poor patient survival rate. Thus, novel side-effect free therapies with higher effectivity are pressingly sought as an alternative to treat cancer patients.

Among the existing alternative treatments, localized hyperthermia has flourished as a remarkable curative modality in which thermal agents are used to transfer heat to the region around cancer cells, promoting their necrosis [4,5]. In particular, nanomaterials-based Photothermal Therapy (PTT) is appealing, since the ability to generate highly focused temperature spots diminishes the damage of surrounding healthy tissues. PTT relies on nano-materials as photothermal agents and near infrared (NIR) laser irradiation to increase their temperature. Unlike conventional cancer therapies, PTT is a minimally-invasive technique that may potentially allow a fast patient recovery with controllable, high selective and favourable biosafety features [6,7].

The success of the PTT relies strongly on the local heating source, i.e., the optically absorbing agent and its biofunctionalization [8], which is required to recognize the carcinogenic cells, for selective targeting. Various types of photothermal agents have been investigated for PTT applications, including bovine serum albumin heterojunctions [9], graphene nanoparticles and carbon nanotubes [10], but the PTT agents based on Localized Surface Plasmon (LSP) resonances are more widespread and promising [11]. These nanoheaters generate heat via resistive losses associated to light absorption at the plasma eigenfrequency. The absorption wavelength is very sensitivity to the nanoparticle shape, size and composition [12–14], hence allowing the tunability of the heating. The ability to control heating at the nanoscale has opened a new research area [15], thermoplasmonics, which encompasses most modern applications of nanoscale heating using plasmonic effects, including PTT.

Gold, among others like platinum, is a widely used material in PTT applications, due to its biocompatibility, weak cytotoxicity, and low reactivity. Most noble metals feature LSP resonances in the UV-Vis range. In particular, gold LSP resonances can be shifted to the first NIR bio-window (NIR-I, 700-900 nm) [16], where light/heat conversion can be maximized due to a minimization of light absorption by water. Photothermal agents in the so-called second NIR biowindow (NIR-II, 1000-1400 nm) offer advantages in clinical applications due to the deeper penetration depth and lower optical absorption compared with NIR-I. Up to now, most of the LSP-based photothermal agents applied for PTT are mainly restricted to single-metal nanoheaters including gold nanospheres [17–20], nanorods [21], nanostars [15,22] or nanodisks [23]. In fact, we have recently compared the thermoplasmonic behaviour of these geometries against doughnut-shaped nanoheaters, showing that the latter feature better thermal performance for photothermal applications [24].

Despite its high heating efficiency, these single-metal nanostructures heat up the surrounding environment isotropically leading to a partially uncontrollable heating effect. This feature does not only carry the potential damage of the targeted cancerous cells, but also the undesirable damage of the neighbouring healthy cells. This is one of the challenges we need to surpass for a successful incorporation of PTT in clinical treatments. Fortunately, the development of novel fabrication methods has provided feasible resources to overcoming this issue, offering novel structures with a superior thermo-optical performance, mainly based on the use of hybrid nanomaterials [25–27].

Janus Nanoparticles (JNPs) are nanostructures composed by separated regions of different chemical compositions. They have attracted great attention in the area of nanomedicine [28,29] because the combination of materials with different chemical and physical properties yields unique dual-functional capabilities offering new opportunities in fields such as imaging or drug delivery [30] for biomedical applications. Colloidal JNPs have also been eventually explored to enhance the thermal performance of heating agents, such as multifunctional ternary JNPs [31], UFO like hybrid cyclodextrin-Pd nanosheets [32], silver-silica nanoplatforms [33], gold triangle-mesoporous silica structures [34], hybrid gold nanostars [35,36] or different core-shell geometries [37–39]. So far, although the application of multifunctional hybrid particles is growing rapidly in nanomedicine, the design of highly efficient single JNPs with exceptional directional photothermal conversion abilities is still missing. Nanoheaters commonly show a remarkable symmetrical temperature distribution (around the main axes of the nanoparticle for anisotropic geometries) hence heating the environment around them homogeneously. To enable heating in specific directions, alternative nanostructures must be designed.

In this work, we present a novel and effective prototype of a Janus heating nanojet (called J-Nanojet from now on) that offers outstanding photothermal conversion features together with unique directional heating capacity, being capable to channel in one direction up to 91% of the total heat. The J-Nanojet is made by a metallic toroidal nanoheater unit embedded in a Janus capsule built of two different biocompatible materials, one acting as heat insulator and the other as heat conductor. We propose the use of fully biocompatible plasmonic materials such as gold or platinum [40] as heating agents to tune the working wavelength of this J-Nanojet from NIR-I to NIR-II depending on the specific needs of applications. Finally, we satisfactorily tested the heating performance of the J-Nanojet under dynamic flow conditions, i.e. considering the thermal behaviour stability to J-Nanojet random rotations when illuminated with unpolarized light, as expected to happen in a practical situation.

## 2. Methods

Noble metal nanoparticles are highly lossy so that heat can be generated under optical excitation, specially at the plasmon resonance. The incident electric field strongly drives mobile carriers of the metal, heating the material owing to the electrons scattering. Then, heat diffuses away from the nanostructure leading a temperature increment of the surroundings. This phenomenon can be described by the classical heat equation in the absence of phase transformations:

$$\rho(r)c(r)\frac{\partial T(r,t)}{\partial t} = \nabla k(r) \nabla T(r,t) + Q(r,t) \qquad (4)$$

where r and t are the position and time, T(r,t) is the local temperature and the material parameters $\boldsymbol{\rho}$(r), c(r) and k(r) are the mass density, specific heat and thermal conductivity, respectively. The function Q(r,t) represents the energy (heat) source coming from electromagnetic losses. Therefore, for a fully characterized system (well-known nanoparticle and surrounding thermo-optical properties), the electromagnetic problem must be solved to obtain these losses. To do so, the system of Maxwell's equations, and subsequently, the heat equation with appropriate boundary conditions have been solved by means of finite element simulations. For convenience and reliability of the solution, we have used Comsol Multiphysics 5.6, which provides state-of-the-art methods for partial differential equations solving. In our simulations, we have assumed that particles are immersed in water and illuminated with circularly polarized light to emulate the more practical unpolarized illumination. Unpolarized light can be simulated by just solving the problem for two orthogonal input electric fields and then taking the mean of the results, but this doubles the simulation time. However, circular polarization gives equivalent results while being computationally more efficient. We have considered the electromagnetic losses as the only heat source and the thermo-optical properties have been taken from Comsol database. To consider heat dissipation in our simulation region, we used a heat flux node across the outer boundaries, considering a heat transfer coefficient, dependent on the geometry and the ambient flow conditions. The heat transfer coefficient h can often be estimated by dividing the thermal conductivity of the fluid by a length scale.

## 3. Results

### 3.1 First steps: Hybrid nanoparticle designs

Most experiments focused on photothermal biomedical applications consider specifically biofunctionalized metallic nanoparticles in a fluid, which eventually binds to the target cells [41]. These nanoparticles, under an adequate illumination, become an LSPR-assisted heating source. In the specific case of PTT against cancer, the local temperature of the tissues must reach 42-48 degrees centigrade [7,15], a temperature increment of approximately 10 ºC above the average temperature of the biological medium.

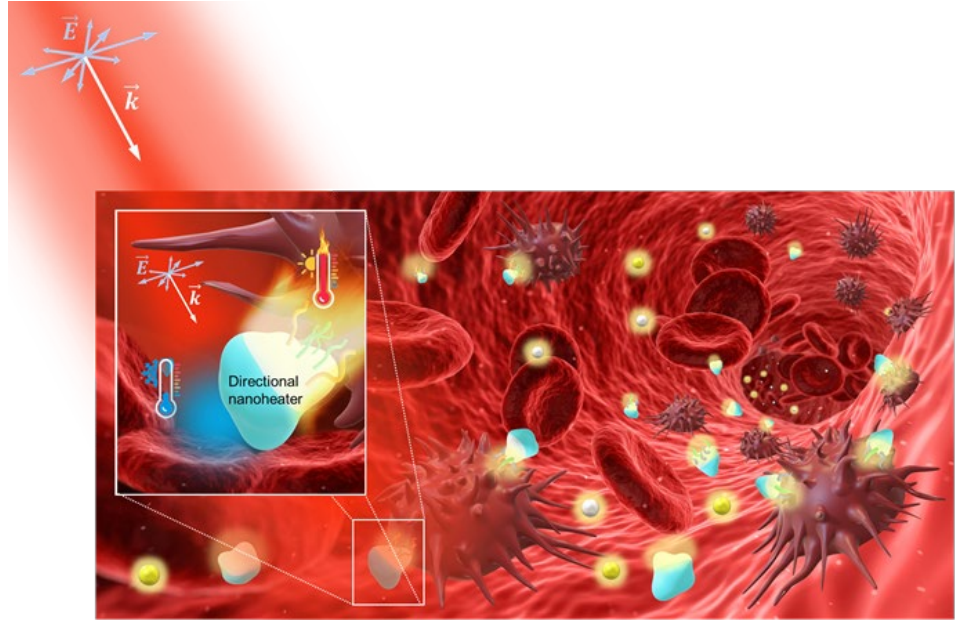

Figure 1. Illustration of spherical nanoparticles and abstract hybrid nanonheaters flowing inside a blood vessel in colloid under infrared laser beam illumination. Inset: abstract J-nanojet structure reacting to the incident light, heating up a target cancerous cell. Picture background: "Creative Commons" by qimono (CC BY-SA 4.0).

As illustrated in figure 1, the temperature spatial distribution generated by single nanoparticles immersed in a fluid has strong symmetries, as is the case of the gold nanospheres, resulting in a radial stationary isotropic thermal flux. This spatial uniformity leads to a dramatic reduction in the heat transfer control since the thermal energy flows in all directions, i.e. only a portion of the generated heat reaches the target cells, with the rest of the heat diffusing towards surrounding tissues. Consequently, a great part of the available thermal energy is wasted. In light of this, to increase the heating efficiency and to maintain its biosafety, it is vital to improve the thermal capabilities of the nanoheaters in terms of temperature generation but more important, in thermal flow control. A clear step forward from the existing single-material structures is to consider asymmetric nanoparticles made of two different materials, a typical JNP, where each half of the particle volume is occupied by one of them, we call this a "hybrid nanoheater". Both materials must be selected with radically opposite thermal conductivities. We propose a hybrid nanoheater composed by polydimethylsiloxane (PDMS) and gold or platinum, acting the first one as an efficient heat insulator and the second one as the heating source.

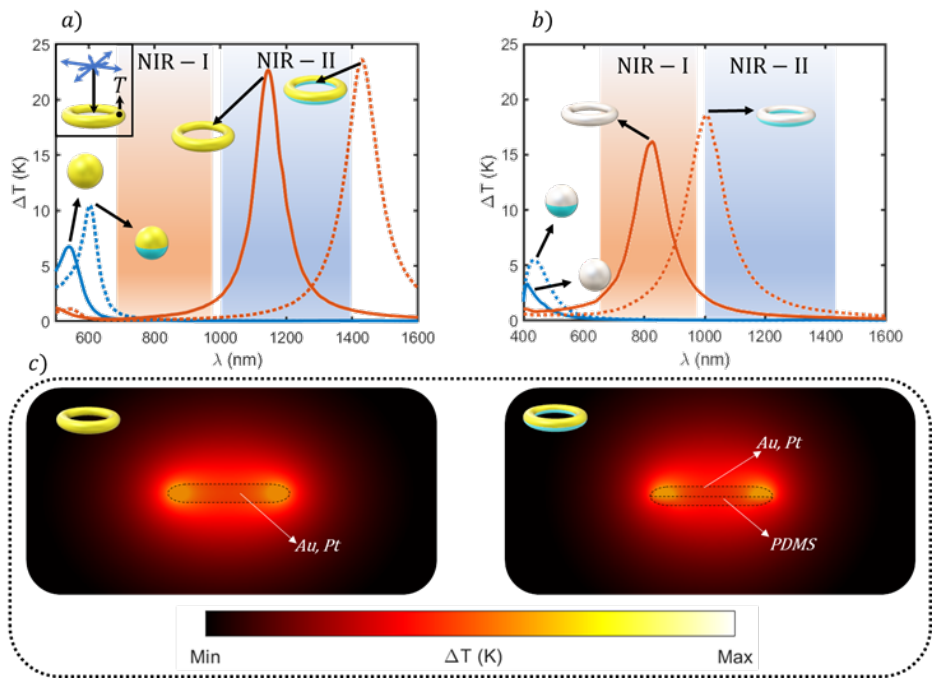

Figure 2. Thermal spectrum comparison of the studied single-material and hybrid nanoheaters for gold (a) and platinum (b) irradiated with a power density of 0.1 mW/μm$^2$; solid and dotted lines correspond to single material and hybrid nanostructures respectively. The reddish and blue shadowed regions represent the first and second biological windows (NIR I-II) respectively. The temperature is taken at the point T, inside the metal material for all structures. c) Temperature maps for the optimal wavelength in (a) for the single material and hybrid toroids immersed in water.

Figure 2, shows a comparison of the heating spectral response of a single material nanoparticle with a hybrid Janus one. Two different shapes are considered: sphere and toroid. In figure 2(a) the particles are made of gold and in (b) made of platinum. The dimensions of the nanoparticles acting as nanoheater units were selected from the optimal ones reported in [24]: The optimal sphere has a radius of 40 nm while the toroid has a 50 nm main radius and 10 nm secondary radius. It can be seen that in the single-material sphere case, the maximum achievable temperature increment ($\Delta T$) increases when the isolating material is added for both, Au and Pt, and their thermal response is also redshifted, approaching the NIR-I. However, both maxima show a poor temperature increment for the fixed incident power density (0.1 mW/$\mu m^2$) and are located out of the NIR biowindow, becoming inadequate for biomedical applications. On the other hand, the toroidal-shaped nanoheaters display superior heating performances with respect to the spherical cases.

It is noteworthy the effect of the particle materials. The spectrum of the Au toroid lies out of the NIR-II when PDMS is added. Although that response can be tuned back to the NIR-II by reducing the dimensions of the toroid, the maximum $\Delta T$ is also decreased (see Fig. S1 of the supplementary material). This dramatically reduces its heating potential and consequently, the applicability in biomedical applications. The Pt toroid spectrum is also redshifted with respect to the single-material nanoheater, remaining at the threshold of the NIR-II. Despite this hybrid nanoparticle offers a favorable temperature increase, particles able to respond to excitation wavelengths further in the NIR-II are more desirable since the tissue penetration length is larger in that region. In Figure 2c the temperature maps of single material and hybrid nanodoughnuts are shown. It can be clearly seen that both architectures exhibit a similar temperature distribution, but the hybrid one is slightly asymmetrical. Thus, the typical Janus nanoparticle, called here hybrid nanoheater, presents a low degree of anisotropy, suggesting a poor achievable directional transfer. These temperature distributions remark the necessity of improving the nanoheater thermal directivity while preserving its power generation efficiency.

However, although some insights can be obtained from the temperature spatial distribution to estimate the thermal directional capabilities of nanoparticles, it is necessary to quantify their anisotropy degree to select the best candidate depending on the specific biological application. This can be achieved by calculating the resulting thermal flux in a fluid. We assume that radiation heating can be neglected regarding the magnitude of the temperatures we are considering (biological medium temperature). The convective heating can also be neglected since surrounding fluid is supposed to be stationary, being the system analysed only half micron far from the nanoheater. Therefore, we are considering that the heat transfer in a complex biological fluid is expected to be mainly owing to conduction [38].

To properly characterize the directional capabilities of nanoheaters, i.e., the ability to drive directionally the heating power, two magnitudes must be analyzed: the relative power flowing through the upper and bottom half-spaces ($Qe$) and the total generated power ($P$). The followed strategy to calculate these magnitudes is illustrated in the inset of figure 3. A cubic surface surrounding the nanoheater and centered at its geometrical center has been considered. This geometry has been selected for convenience in the data management of the simulation software (Comsol Multiphysics). Then, the temperature gradient is integrated along the upper ($S_u$) and bottom ($S_b$) half-spaces (integration domain) to obtain the ratio of heating power going through them. The results of the estimated relative heat power for each nanoheater flowing through the upper media are shown in Figure 3.

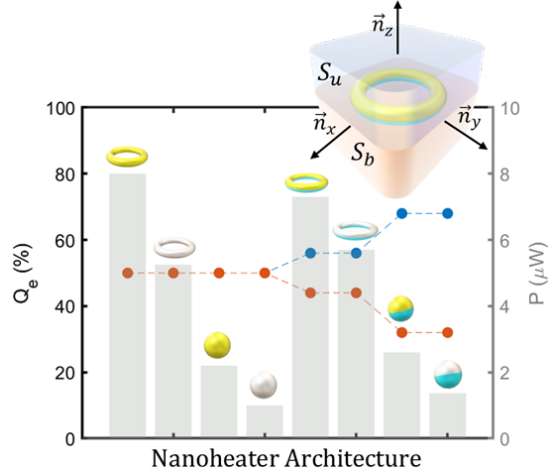

Figure 3. Comparative of the relative conductive heat power ($Qe$) flowing through the upper half-space (in blue) and lower half-space (in orange) for single material and hybrid nanoparticles irradiated with a power density of 0.1 mW/µm$^2$ at the resonance. The grey bars correspond to the generated total heating power for each architecture. Inset: Scheme of the conductive thermal power calculation method applied to the nanoheaters where $S_u$ y $S_b$ are the upper and bottom half-spaces respectively. $n_x$, $n_y$ and $n_z$ are the canonical basis vectors.

It can be seen that spherical nanostructures reach a maximum value of 68%, while toroidal nanostructures offer 56%, in contrast with the single-material particles which as expected, are totally symmetric displaying a relative integrated flux of 50%. Nevertheless, although hybrid spheres display higher directional capabilities, they generate the lowest thermal powers when compared to toroidal geometries. Therefore, the hybrid nanoheater geometries clearly show an improved heat directional transfer with respect to the single-material structures while keeping the achievable maximum temperature (figure 2) and the generated total power (figure 3) almost constant. Even though this is already an interesting result, we take this one step further to magnify both, the directional capabilities and the total generated heating power for each architecture.

*3.2 J-Nanojet design: Thermal response optimization*

In sight of the previous results, designs that allow to enhance the heating performance asymmetry while improving the maximum achievable temperature and power are sought. For these purposes, we have designed and optimized novel structures capable of act as heating nanojets, propelling in a highly directional manner most of the thermal energy generated by the nanoheater unit, while reducing the heating of the opposite side.

The illustrations in figure 4(a-b), show the proposed basic scheme of the two different J-Nanojet designs, so-called binary and ternary. A cubic geometry has been considered for convenience in its implementation in the Comsol software. The first design (figure 4a) consists of a doughnut-shaped metallic nanoparticle embedded in a Polydimethylsiloxane (PDMS) capsule (binary J-nanojet), touching the upper boundary of the polymer in contact with the surrounding fluid. The second one (shown in figure 4b), is an extension of the binary nanojet, by just adding a thin diamond cap on top of the toroid to force the heat to flow in the upper direction (ternary J-nanojet). PDMS is an effective thermal insulator and diamond a good thermal conductor, showing thermal conductivities of ~ 0.15 W/mK [42] and 2200 W/mK, respectively [43]. Furthermore, considering its good thermal performance, the optimal toroid from [24] has been selected as the heating unit. This selection guarantees the maximum achievable temperature increase for excitation wavelengths in the NIR biowindow. This is essential in real PTT applications where nanostructures are randomly oriented and human tissues induce a partial depolarization of light with an exponential decay of its intensity. Hence, the J-Nanojets are always analyzed under unpolarized illumination with a pumping power density of 0.1 mW/μm$^2$. Furthermore, in PTT, nanoheaters would circulate through thin capillaries, thus, the size of the agent acting as nanoheater must be controlled and below a certain allowed value. This condition has been satisfied considering nanojet designs with all dimensions around a therapeutically allowed of 150 nm [12].

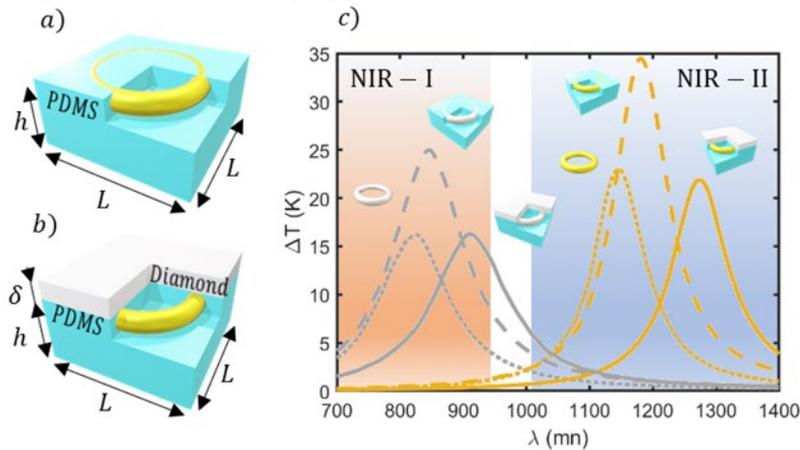

Figure 4. Illustration of the J-Nanojet architectures: a metallic toroid embedded in a PDMS block (a), covered with a thin squared diamond layer (b). An illustrative section has been made in the nanoparticles to visualize the metallic heating unit. The variables *h*, *δ* and *L* represent the capsule parameters: PDMS thickness, diamond thickness and the size of the capsule base respectively. c) Thermal spectrum comparison of the structures shown in (a) and (b): an isolated metallic toroid (dotted line), a toroid embedded in a 125x125x50 nm block of PDMS (dashed line) and the ternary J-nanojet structure (solid line). The blue and reddish shadowed regions represent the first and second biological windows respectively.

A previous optimization of the capsule geometrical parameters has been performed to visualize their effect in the J-nanojet thermal response. Different values of the conductor and insulator thicknesses ($\delta$, $h$) were considered, ranging from 10 to 30 nm and 20 to 50 nm, respectively (see Figure S2 in the supplementary material), reporting a notorious maximum temperature increase stabilization for $\delta = 10$ nm and $h \geq 50$ nm. The block-base ($L$) of the nanojet has been considered to be 125 nm side to allow the torus to be fully embedded in the PDMS block while allowing for compactness. Figure 4c), shows the results obtained after that optimization, comparing the J-nanojet designs with the isolated heating unit. This allows to select the best design in terms of potential heating capability, i.e., maximum temperature increase and spectral location of the thermal response. Notice also that, although similar trends are found for both heating unit materials (Au and Pt) in both J-nanojets (binary and ternary), the platinum thermal response is broader, thermally weaker, and its effective response is located at the NIR-I.

On the other hand, the gold core gives higher temperatures but at the NIR-II. This fact opens the possibility to extend the use of this nanostructure to other biomedical applications, where the tissue penetration length is not a critical parameter and gives the platinum structure more tolerance to fabrication imperfections. However, it can be clearly seen that in terms of the maximum temperature achievable, the binary design outperforms the ternary one, reaching increments of 53-58% while equivalent values are obtained compared with the isolated doughnut (see figure S3 to visualize the thermal response of the ternary design covered with sapphire as an alternative to diamond).

For a better understanding of the J-Nanojet thermal response and to show why they can suppose a significant advance in certain photothermal applications, such as PTT, the previous results are complemented with their spatial temperature distributions. Before analyzing the spatial thermal distribution that these nanojets can offer, an illustration of the working principle is shown in figure 5a). As can be seen, the heat flows preferentially through the material with higher thermal conductivity, while PDMS introduces a resistance to heat flow. The heterogeneous structure of the material induces a noticeable asymmetry in the temperature profiles along the z axis (binary nanojet), which can be enhanced by adding a thin diamond layer (ternary nanojet). Thus, the asymmetrical z-profiles suggest that both architectures may present a high temperature contrast between the front (A for ternary and B for binary J-nanojets) and rear boundaries (C). This induced thermal Janus effect, enables the possibility of affecting the target cells while protecting the rest of healthy tissue.

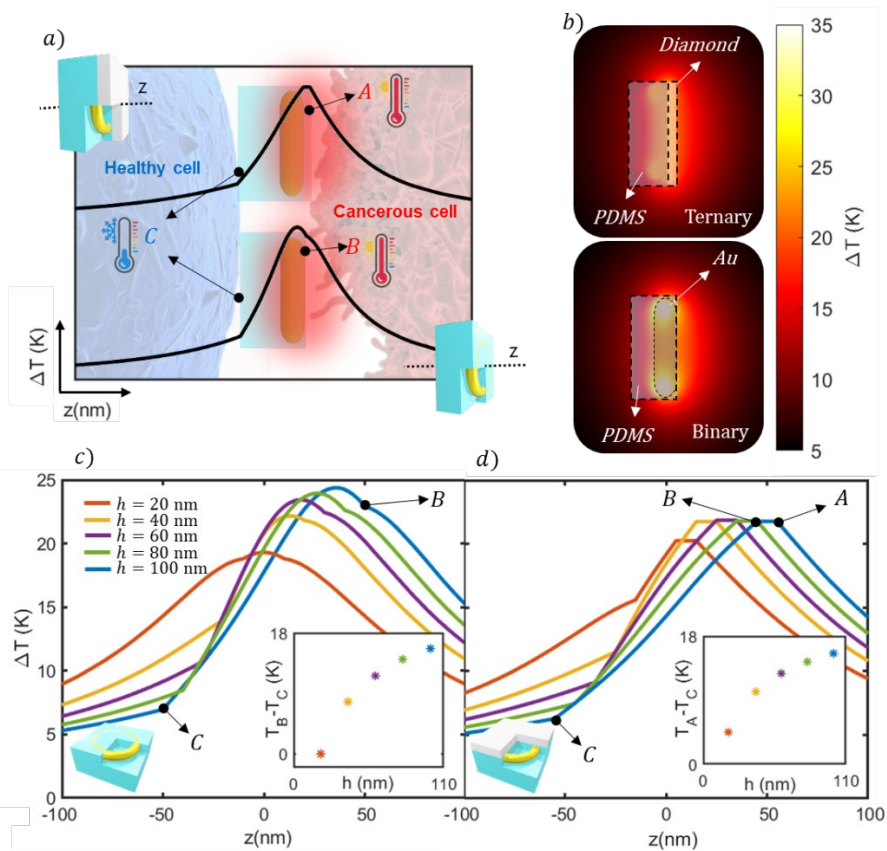

Figure 5. a) Illustration of the typical thermal profile taken along the *z* axis for the J-nanojet architectures where A and B are the PDMS boundaries, and C is the boundary of the diamond. b) Temperature spatial distribution of the different J-nanojets in water excited at $\lambda$=1180 nm (binary) and $\lambda$=1275 nm (ternary). Comparison of thermal profiles along the *z*-axis for a set of insulator thicknesses ranging from 20 to 100 nm in 20 nm steps for the binary (c) and ternary (d) J-nanojets.

Figure 5b shows the temperature maps for the optimal binary and ternary J-nanoheaters, where the aforementioned thermal distribution asymmetry is demonstrated. Although the binary nanojet presents the highest temperature increase, the corresponding temperature map evidences how this temperature is confined inside thermal insulator, resulting in heat inefficiently transmitted to the surrounding fluid. Furthermore, both J-nanojets show comparable spatial thermal distributions, that tend to be almost isotropic at large distances from the heating device. However, in the case of PTT, the tumor cells would be located at a distance of around 10-30 nm from the J-Nanojet surface due to the biofuntionalization agents [44] (see Figure 1), so that the short distance heating performance needs to be analyzed in detail. Moreover, one of the most important aspects to consider is the precision in the thermal control. In photothermal applications, the temperature increase is sought to be stable and controllable, so temperature gradient-free structures are desirable since they allow for a high temperature decay control along a great surrounding volume. In order to get deeper insights of an optimal thermal agent prototype, it is necessary to analyze how the temperature decays with respect to the distance from the front and back nanoheater surfaces.

Figures 5c-d show the thermal profiles of the binary and ternary J-nanojet excited at their most efficient wavelengths, 1180 nm and 1275 nm, respectively (extracted from Figure 4c). It can be seen that the binary design shows higher temperature increments than the ternary, being comparable for thinner insulator layers. Equivalent results were obtained for the platinum heating unit (see Figure S4 in the supplementary document). An interesting situation appears for the 20 nm-thickness case, where it can be seen that the binary design offers a totally symmetric z-profile, leading to an absence of conductive thermal flow control, while the ternary nanojet presents an asymmetrical profile from the beginning, allowing a certain heat directionality. In that case, the temperature contrast between the front and the rear of the nanoheater side is converted from 0 to 4.6 K. Thus, the ternary nanojet architecture offers the possibility to partially propel forward the heat power, highly reducing backward heating, even for smaller insulator coverage which is favorable given the size limitations inherent to the biological medium of interest.

In regard to Figure 5c-d, it is also noticed how the maximum temperature reached along z-axis by both architectures evolves with the insulator size. In the binary design case, this temperature increases as the PDMS is added, being stabilized for a thickness of around 100 nm. Conversely, the nanojet shows a remarkably stable temperature increment for a wide range of PDMS thicknesses, remaining its optimal thermal response intact facing fabrication imprecisions. Therefore, the nanojet structure may offer high tolerance to fabrication imprecisions being capable to direct heat for all thicknesses. On the other hand, the thermal response of the binary design is more sensitive to thickness variations becoming less tolerant to manufacturing irregularities. However, it presents a higher maximum temperature becoming more effective in light to heat conversion, so that, both structures offer an outstanding thermal performance regarding photothermal applications.

A highly influential parameter to consider in the heating control optimization is the temperature gradient between the front and the rear of the nanoheater since it gives a clear idea about the nanoparticle ability to heat up only in forward direction, insulating backwards. The tendency of this magnitude is shown in the insets of figure 5c-d. For all cases, as the insulator thickness increases, the temperature contrast grows, leading to a profile asymmetry enlargement. Consequently, both architectures are more effective directing heat for higher insulator thicknesses, disregarding the core material.

*3.3 J-Nanojet directionality: Thermal conductive flux*

In light of the previous results, the amount of heating power that J-nanojets can provide and properly drive is calculated. The approach is similar to that followed for the hybrid structures. As illustrated in the inset of Figure 6 a), the heat flow through the faces of a cubic surface with the J-nanojet at its centre is analyzed. The results are shown in figure 6 a), where the obtained heating power for binary and ternary J-nanojets can be seen. One interesting feature that distinguishes both architectures is the relative heat power discrepancy for thin insulator thicknesses. As seen in the figure, the ternary J-nanojet offers certain heating power control even for the thinnest PDMS covering.

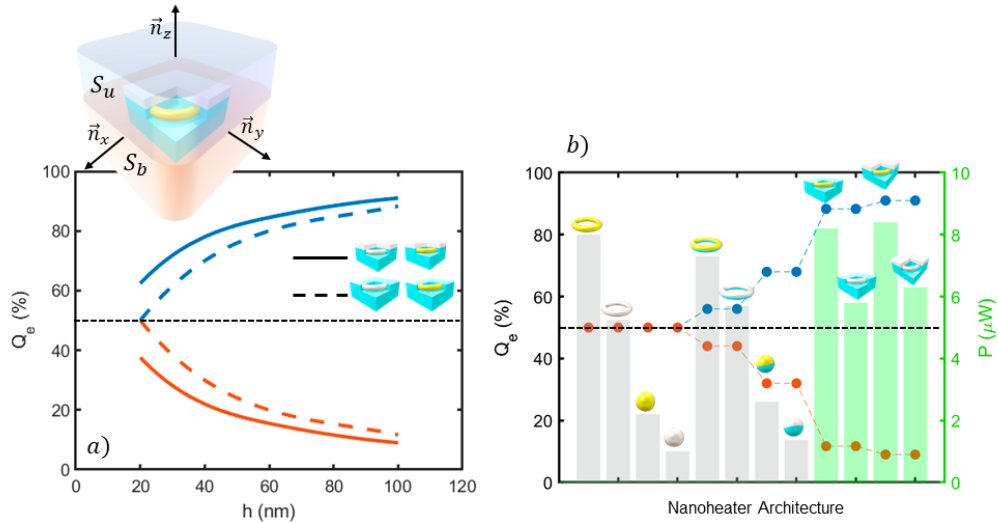

Figure 6. a) Comparative of the relative conductive heat power ($Q_e$) flowing through the upper half-space (in blue) and lower half-space (in orange) as a function of the insulator thickness $h$ for gold and platinum heating units and both, binary (dashed line) and ternary nanojet (solid line) designs. Inset: Scheme of the conductive thermal power calculation method applied to the nanojets where $S_u$ y $S_b$ are the upper and bottom half-spaces respectively. $n_x$, $n_y$ and $n_z$ are the vectors of the canonical basis. b) Comparative of the maximum relative conductive power for all the considered structures: single-material/hybrid nanoheaters in grey and J-nanojet designs in green.

Furthermore, as PDMS is added, the relative heating power directivity increases for both architectures since the insulator material inhibits heat transport through it while the diamond cap in the ternary nanojet promotes a flow in forward direction. Consequently, the efficacy of the binary and ternary nanojet structures are closer for larger PDMS sizes being able to direct forwards up to the 87% and 91% of the heating power respectively. These high directivities outperform the obtained for hybrid nanoparticles as illustrated in Figure 6 b), where a maximum value of 68% is reached for spherical nanoheaters. The toroidal geometries offer a directivity of 56%, in contrast with the single-material particles which are totally isotropic. It must be noticed that under a 0.1 mW/µm² power density illumination, spherical nanoparticles show the lowest generated thermal powers for all cases, becoming inferior compared with toroidal geometries. In contrast, both J-nanojets designs generate the highest thermal powers reaching more than 8 µW and 6 µW for gold and platinum cores, respectively. Thus, the J-nanojet designs are capable of generate more heating power for similar input conditions resulting in more efficient architectures that are able to propel the greatest part of that energy forwards.

*3.4 J-Nanojet in fluid: Thermal dependence with orientation*

As reported in [24, 45], another remarkable and highly influential aspect to consider in JNP-assisted biomedical applications is the fact that nanoparticles are immersed in a flowing free biological environment, it makes the nanoparticles to dynamically suffer rotations that strongly affect their photothermal conversion efficacy. Therefore, the effective temperature obtained in such a case is the average of temperatures achieved for all possible orientations with respect to the electric field of the incident light being more favorable those particles with a poor thermal dependence on the electric field orientation. Furthermore, biological tissues are expected to be absorbing and dispersive, properties that dramatically reduces the incident light degree of polarization.

To clearly discern the stability of the J-nanojets under free rotations, we considered a ternary J-nanojet immersed in water and calculated how their relative orientation to the incident light direction affects their thermal response. To do so, the thermal response of the ternary J-nanojet was calculated for the optimal wavelengths taken from figure 4c for a set of 3D rotations. Then, the numerical mean temperature increment is calculated. Figure 7 shows the temperature maps of the nanoparticles for all the studied 3D rotations with respect to the $x$ and $y$ axes, taking their optimal response orientation as the initial position. The system is analyzed for unpolarized light.

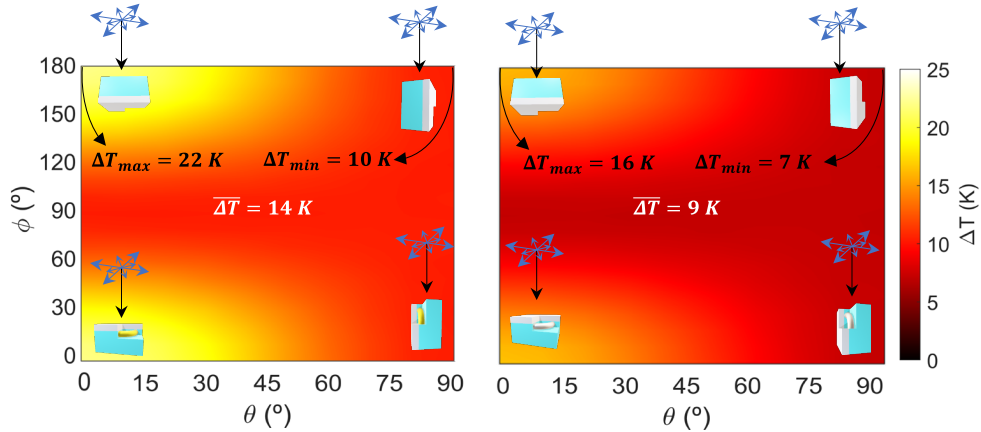

Figure 7. Thermal responses of the gold and platinum ternary nanojet to rotations. $\theta$ and $\phi$ are the rotations with respect to the $x$ and $y$ axes. The considered unpolarized light (insets) injects a power density of 0.1 mW/μm$^2$.

Attending to the Figure 7, it can be clearly seen a revolution symmetry in the colormap, as the revolution axis of the nanojet and the $z$ axis (beam direction) match in the initial configuration. Thus, considering the temporal evolution of the electric field vector for unpolarized light, rotating the nanojet about the $y$ axis, and subsequently about the $x$ axis, has the same effect as rotating it about the $x$ axis and then about the $y$ axis. To clarify this figure, setting $\theta = 0°$ and then increasing the rotation in $\phi$ will mean that eventually the transverse length of the toroidal metallic core and the electric field will be orthogonal, leading to occasional resonances due to alignments between this transverse length and the rotating electric field, resulting in a fall in temperature increment around 53% and 56% compared to the most favorable configurations ($\Delta T=22$ K and $\Delta T=16$ K) for gold and platinum, respectively. Thus, due to the nanodoughnut symmetry, the nanojet exhibits a more stable thermal response to rotation for both materials, offering average temperature increments of 14 K and 9 K, accordingly (a similar investigation for different capsule geometries can be found in Figure S5).

## 4. Conclusions

We have performed a numerical investigation of the photothermal response of different Janus nanostructures, showing that a combination of materials with radically different thermal properties (PDMS/diamond) produces a strong thermal Janus effect allowing to control the heating direction and improving the efficiency of nanoheaters. The proposed doughnut-based architectures, the binary and the ternary nanojet designs, outperform the simple Janus nanostructures such as the sphere and toroid geometries in terms of maximum temperature increase and thermal control, becoming a powerful tool in biomedical applications since it provides larger temperature areas supporting resonances within the therapeutic windows and also reducing the backward heating. On the one hand, the binary design displays a superior maximum temperature increase, however, it is more sensitive to insulator thickness variations, resulting in a lower tolerance to fabrications imperfections. Meanwhile, although the hybrid nanojet structure offers a slightly weaker thermal response, it provides higher thermal control together with a remarkable tolerance to insulator thickness imperfections. We have also analyzed the impact of hybrid nanojet rotational diffusion on its photothermal performance. This structure features a high average temperature increment also showing a great tolerance to rotation which makes it an ideal candidate for biomedical applications.

## Conflicts of interest

There are no conflicts to declare.

## Acknowledgements

Authors would like to thank Profs. O. Muskens and C. R. Crick for the interesting and valuable discussions. We gratefully acknowledge financial support from Spanish national project INMUNOTERMO (No. PGC2018-096649-B-I), the UK Leverhulme Turst (Grant No. RPG-2018-384), UK-EPSRC (EP/J003859/1) and Imperial College Europeans Partner Fund grant. J. G-C. thanks the Ministry of science of Spain for his FPI grant. G. S. thanks the Ministry of education for his collaboration grant and P.A. acknowledges funding for a Ramon y Cajal Fellowship (Grant No. RYC-2016-20831).

# Supplementary Information

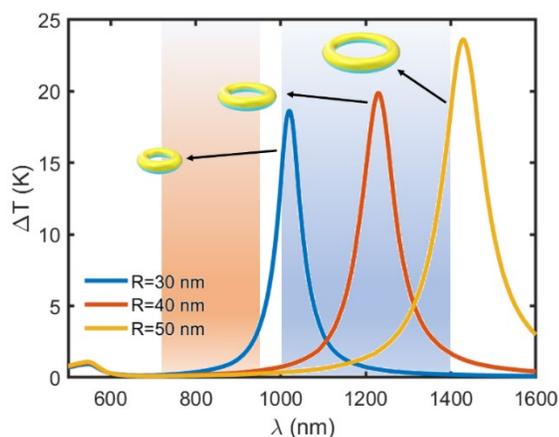

*Figure S1. Thermal spectrum comparison of different Au/PDMS hybrid nanodoughnuts with 30, 40 and 50 nm main radii and 10 nm secondary radius. The reddish and blue areas determine the NIR-I and NIR-II respectively.*

Figure S1 shows the thermal spectral response of hybrid nanodoughnuts composed by PDMS and gold for different main radii. As can be expected, when this parameter decreases the maximum temperature increment is reduced and so does the resonance wavelength. This leads to a compromise between size and thermal performance of nanoheaters. In this work, the best nanoparticle in terms of maximum temperature achievable has been considered, being selected the 50 nm main radius toroid.

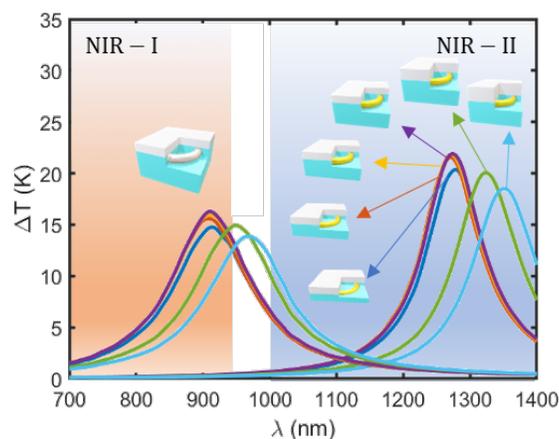

*Figure S2. Capsule spectral thermal optimization for binary and ternary designs. The PDMS thickness varies from 20 to 100 nm and the diamond one from 10 to 30 nm.*

Figure S2 shows the spectral thermal response of the ternary J-nanojet for different conductor/insulator thicknesses. The insulator thickness varies from 20 (which is required for the toroid to be fully embedded in the insulator block) to 100 nm and the conductor one from 10 to 30 nm. It can be seen the expected redshift of the platinum toroid response with respect to the gold case. In sight of the figure, the diamond cap has the major impact in the spectral location due to its high refringence. The maximum temperature is also decreased as diamond is added by cause of the thermal conductivity (2200 W/mK). Thus, the optimum spectral response is reached for insulator thicknesses between 50 and 100 nm and for a conductor thickness of 10nm.

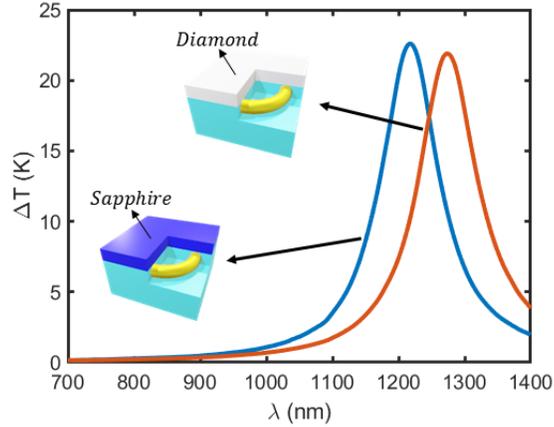

*Figure S3. Comparison of the thermal spectrum for the nanojet structure covered with a 10 nm-thick layer of sapphire (blue) and diamond (red). The PDMS thickness is 100 nm.*

Figure S3 plots the thermal response of two nanojets covered with a 10 nm-thick layer of sapphire and diamond respectively. It can be clearly seen how the diamond produces a spectral redshift with respect to the sapphire one. This can be analysed considering the refractive index values for both materials. In this spectral region, diamond presents a refractive index of around 2.38 RIU while sapphire possess 1.75 RIU. However, the most remarkable aspect is the slight discrepancy in the temperature increment. Despite the huge contrast of thermal conductivities (2200 W/mK for diamond and 35 W/mK for sapphire), the maximum temperature increase remains similar for both cover materials. This suggest that, disregarding the impact in the temperature increment, different materials with a wide range of thermal conductivities can be selected for the conductive cover layer, only considering other aspects such as the fabrication possibilities.

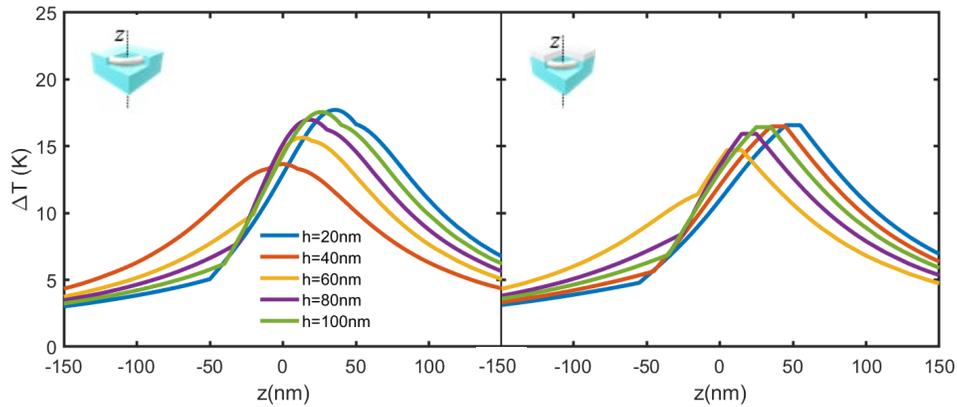

*Figure S4. Extended comparison of thermal profiles along the z-axis shown in Figure 4 for a set of insulator thicknesses ranging from 20 to 100 nm for the platinum toroidal core.*

Figure S4 shows the thermal profiles along the z-axis for the platinum core (binary and ternary J-nanojets) Accordingly to the figure 5 in the manuscript, it can be seen that the presence of a diamond layer in the ternary design induces a thermal stabilisation in its volume where the maximum temperature is reached. In contrast, in the binary design this maximum is reached inside the PDMS becoming useless. As in figure 5, paying attention to the thinnest PDMS capsule ($h$=20 nm), it can be seen that the ternary design offers certain asymmetry from the beginning while the binary does not. This can be favourable regarding the size limits required in biological media. Furthermore, the same behaviour of gold can be observed. As the PDMS thickness grows the maximum achievable temperature increases, being immediately stabilised in the ternary J-nanojet.

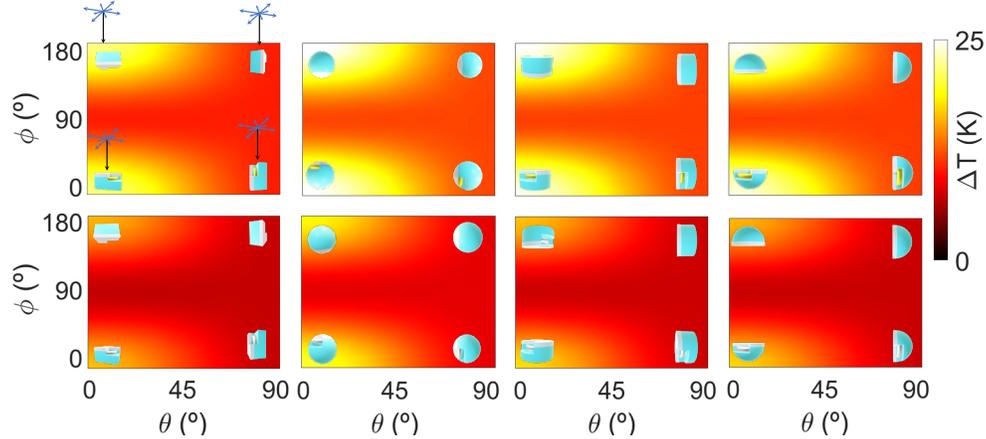

*Figure S5. Comparison of the thermal response behaviour under rotations for different capsule geometries: rectangular (a, e), spherical (b, f), cylindrical (c, g), and hemispherical (d, h). First row plots correspond to the toroidal gold core and the second row to the platinum one. The incident beam was unpolarized for all cases.*

Figure S5 shows the angular thermal response for different capsule geometries (rectangular, spherical, cylindrical and hemispherical). The particle rotations have been performed around two main axis, x and y, due to the usefulness of the FEM software as in [19], where an illustration of the rotation system can be found. It can be clearly seen that all the capsule geometries offer equivalent angular responses for both core materials, gold and platinum, showing high stability under rotations in all cases. However, it is noticed that the spherical coverture presents a slight superiority with respect to the others.